\documentclass[prb, superscriptaddress, twocolumn,amsmath,amssymb]{revtex4}
\usepackage{amsmath,amssymb}
\usepackage{tabularx}% Include figure files
\usepackage{graphicx}% Include figure files \usepackage{dcolumn}% Align table columns on decimal point
\usepackage{bmpsize}
\usepackage{bm}% bold math
\usepackage{color}
\DeclareMathAlphabet{\mathpzc}{OT1}{pzc}{m}{it}
\newcommand{\nn}{\nonumber}

\begin{document}

\preprint{APS/123-QED}

\title{Topological Crystalline Superconductivity in Dirac Semimetal Phase of Iron-based Superconductors}

\author{Takuto Kawakami}
\email{t.kawakami@qp.phys.sci.osaka-u.ac.jp}
\affiliation{Department of Physics, Osaka University, Osaka 560-0043, Japan}
\affiliation{Yukawa Institute for Theoretical Physics, Kyoto University, Kyoto 606-8502, Japan}
\author{Masatoshi Sato}
\email{msato@yukawa.kyoto-u.ac.jp}
\affiliation{Yukawa Institute for Theoretical Physics, Kyoto University, Kyoto 606-8502, Japan}

\date{\today}

\begin{abstract}
In iron-based superconductors, band inversion of $d$- and $p$-orbitals yields Dirac semimetallic states.
We theoretically investigate their topological properties in normal and superconducting phases, 
based on the tight-binding model involving full symmetry of the materials.
We demonstrate that a Cooper pair between electrons with $d$- and $p$-orbitals relevant to the band structure 
yields odd-parity superconductivity.
Moreover, we present the typical surface states by solving the Bogoliubov--de Gennes equation and characterize them by topological invariants defined with crystal symmetry.
It is found that there appear various types of Majorana fermions such as surface flat band, Majorana quartet and M\"{o}bius twisted surface state.
Our theoretical results show that iron-based superconductors are promising platforms to realize rich topological crystalline phases. 
\end{abstract}

%\pacs{
%	03.65.Vf %Topological phases (quantum mechanics)
%	42.70.Qs %Photonic band gap materials
%	73.20.At %Surface states, band structure, electron density of states
%}

\maketitle
\section{Introduction}
 Superconductivity in topological insulators~\cite{hasan2010, Qi2011, Andoreview} 
and semimetals~\cite{armitage2018} has attracted considerable attention recently, 
because it can potentially achieve topological superconductivity 
hosting Majorana quasiparticle zero-energy excitation~\cite{TSN2012, Ando-Fu2015, mizushimareview2016, Sato-Fujimoto2016, sato2017, kobayashi2015, hashimoto2016, fu2010, sasaki2011, Fu2014, Bo2015, HeWang2016, matano2016, yonezawa2017}.
Topological insulators and semimetals are realized via band inversion of 
opposite-parity bands at the time-reversal invariant momenta~\cite{Fu2007, qi2008}.
These two bands strongly mix with each other at generic points in the Brillouin zone. 
If the mixed band forms the Fermi surfaces under carrier doping, 
Cooper pairs between opposite-parity electrons are possible.
This type of pairing supports topological superconductivity~\cite{sato2009, fu2010, kobayashi2015, hashimoto2016, sato2017}.

A recent important progress in the search for topological superconductivity is 
theoretical prediction and experimental observation of topological states in iron-based superconductors~\cite{hao2014, zwang2015, hao2015, yin2015, wu2016, pzhang2018sc, pzhang2018NM, gang2017, dwang2018, liu2018, fang2017, machida2018, qin2019, konig2019, kong2019, wang2019, zhang2019, wu2019}. 
In the normal state of these materials, band inversions between the $p_z$-orbital of $p$-block elements and three $d$-orbitals of iron can occur.
The band inversion with one of three $d$-orbitals yields the topological insulating gap.
A recent angle-resolved photoemission spectroscopy experiment with high-energy resolution~\cite{pzhang2018sc} has detected a characteristic surface Dirac cone in an iron chalcogenide Fe(Se,Te).
Even more importantly, this material exhibits superconductivity with a relatively high transition temperature $T_c=13 K$.
Moreover, a superconducting gap on the Dirac surface states~\cite{pzhang2018sc} 
and an energetically isolated zero-energy density of states in the vortex cores
in Fe(Se, Te)~\cite{dwang2018, machida2018, kong2019} and (Li, Fe)OHFeSe~\cite{liu2018}
have been experimentally reported.
The band inversions associated with the other $d$ orbitals yield topological Dirac semimetallic state
as observed in Fe(Se, Te) and Li(Fe, Co)As~\cite{pzhang2018NM}.
The authors indicated the possible topological superconductivity caused by this Dirac semimetal
in collaboration with an experimental group~\cite{pzhang2018NM}, 
although further details of full crystalline symmetry were not presented.

The purpose of this study is to clarify the topological property of possible odd-parity superconductivity 
of the Dirac semimetalic state in iron-based superconductors.  
The crystalline symmetry generally plays important roles to define the topological invariants.
In particular, the symmetry of iron-based superconductor is the non-symmorphic space group $P4/nmm$~\cite{cvetkovic2013}.
The topological states protected by non-symmorphic symmetry have a new class of surface states, 
called M\"obius twisted surface states~\cite{shiozaki2015,shiozaki2016} or hourglass fermion~\cite{zwang2016nat, ma2017}. 
However, this surface state has not been observed yet in superconducting states.

In this study, we develop a theory of topological crystalline phases
realized in normal and superconducting states of iron-based superconductors.
We first construct the simplest tight-binding model describing the topological Dirac semimetal of these materials.
Subsequently, we summarize all the possible Cooper pairs
between electrons with opposite parity residing at the iron and $p$-block element sites 
and classify them in terms of the space group $P4/nmm$.
Furthermore, we clarify the topological invariant and surface states 
depending on the irreducible representation of the gap function and direction of surfaces.
Through our study, we suggest that iron-based superconductors are promising platforms 
to realize rich topological structures protected by their crystal symmetry. 

\section{Normal state}\label{sec:normal}
Let us construct the minimal tight-binding model describing the Dirac semimetal in iron-based superconductors. 
Although our model can be applied to various iron-based superconducting materials, 
we consider the iron chalcogenide Fe(Se,Te).

\begin{figure}[t]
\begin{center}
	\includegraphics[width=85mm]{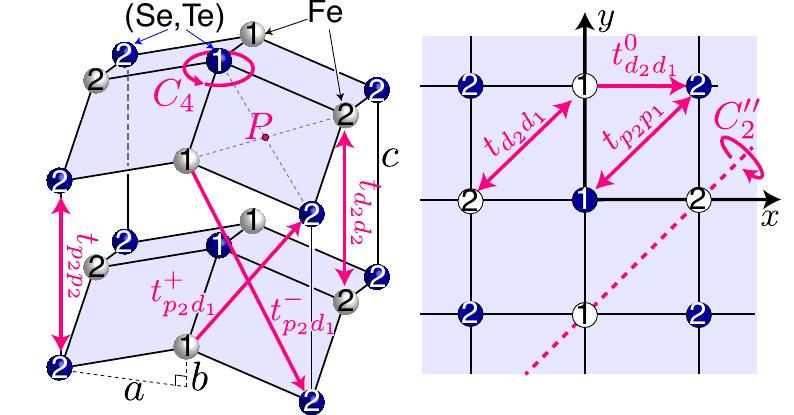}
	\caption{
	Lattice structure of Fe(Se,Te) system. 
	The white and blue spheres indicate the Fe and chalcogen (Se,Te) atoms, respectively,
	with sublattices ``1'' and ``2'' owing to the buckling of the chalcogens.
	}\label{fig:lattice}
\end{center}
\end{figure}

As shown in Fig.~\ref{fig:lattice}, while iron atoms form a flat square lattice,  
chalcogen atoms are displaced in the $\pm \hat{\bm z}$ direction from the iron plane alternately.
Therefore, a unit cell involves four atoms composed of two irons and two chalcogens.
Hereafter, we denote iron atoms as $\alpha=d_1$ and $d_2$ and chalcogen atoms as $p_1$ and $p_2$ 
referring to their outermost orbitals $d$ and $p$.
We take the coordinate such that the iron sites are at $\tilde{\bm r}_{d_1}=a\hat{\bm{y}}$ and $\tilde{\bm r}_{d_2}=a\hat{\bm{x}}$ 
and chalcogens are at $\tilde{\bm r}_{p_1}=b\hat{\bm{z}}$ and $\tilde{\bm r}_{p_2}=a\hat{\bm x} +a \hat{\bm y} -b\hat{\bm{z}}$. 
See Fig.~\ref{fig:lattice}. 

In addition, in the Fe$^{2+}$, Se$^{2-}$, and Te$^{2-}$ of Fe(Se,Te), $3d$, $4p$, and $5p$ orbitals are almost filled, 
and hence, the relatively higher-energy orbitals contribute to the physics around the Fermi level.
First, we focus on the $d$-orbitals of iron.
As shown in Fig.~\ref{fig:lattice}, the neighboring sites of iron are located at the $\pm\bm{x}\pm\bm{y}$ directions. 
In this case, $d_{yz}$, $d_{zx}$, and $d_{x^2-y^2}$ orbitals forming the in-plane $\pi$ bond
have higher energy than $d_{3z^2-r^2}$ and $d_{xy}$ orbitals forming the $\sigma$ bond.
In the absence of spin-orbit coupling, the $d_{yz}$ and $d_{zx}$ orbitals are degenerate 
owing to improper fourfold rotation symmetry around the iron site.
Including the spin degrees of freedom, we have a fourfold degeneracy of $d$-orbitals. 
If we consider the spin-orbit coupling, this fourfold degeneracy splits into 
two sets of twofold degeneracies of 
$[id_{zx}\!\pm \! d_{yz}]| \!\pm\!\frac{1}{2}\rangle$ and 
$[id_{zx}\!\mp \! d_{yz}]| \!\pm\!\frac{1}{2}\rangle$, 
where $|\!\pm\!\frac{1}{2}\rangle$ represents eigenstates of 
up and down spin. 
The $id_{zx}\pm d_{yz}$ orbitals with angular momentum $L_z=\mp 1$ have
the orbital magnetic moment along $\pm \hat{\bm z}$ owing to the negative charge of the electron.
Therefore, the $[id_{zx}\!\mp\!d_{yz}]|\!\pm\!\frac{1}{2}\rangle$ 
($[id_{zx}\!\pm\!d_{yz}]|\!\pm\! \frac{1}{2}\rangle$) states 
where the spin and orbital magnetic moments are antiparallel (parallel)
split to the higher (lower) energy. 
Therefore,  
\begin{eqnarray}\label{eq:d}
\left|d_{1,2},\pm \right>=[id_{zx}- s d_{yz}]\left| \tfrac{s}{2} \right>
\end{eqnarray}
states have the highest energy in five $d$-orbitals at two iron sites.
Here, $s=\pm 1$ is the label of spin.

Subsequently, we consider the $p$-orbital of chalcogen atoms. 
As the iron-based superconductors have a layered structure, the state,
\begin{eqnarray}\label{eq:p}
	\left|p_{1,2},s \right>= p_z \left| \tfrac{s}{2}\right>,
\end{eqnarray}
forming the $\pi$ bond in the $xy$ plane has higher energy than the $p_x$ and $p_y$ orbitals 
forming the $\sigma$ bond. 
According to the first-principle calculation~\cite{zwang2015}, in FeSe, 
this $p_z$-orbital has much higher energy than the $d$-orbitals. 
By partially substituting Se by Te, one can lower this energy to those of the $d$-orbitals. 
This substitution also enhances the dispersion of the $p_z$-orbital along the $k_z$ direction.
Eventually, the $p_z$-band of Eq.~(\ref{eq:p}) intersects the $d$-bands of Eq.~(\ref{eq:d}) between $\Gamma$ and $Z$ points.

To describe this band inversion, we construct the following tight-binding model 
considering the states in Eqs.~(\ref{eq:d}) and (\ref{eq:p}) on each atomic site.
In total, we have eight internal degrees of freedom (d.o.f.)
composed of the two species of atoms with different orbitals, two sublattices, and two spin components.
In general, the Hamiltonian is 
%%%%%%%%%%%
\begin{eqnarray}\label{eq:tb}
	H_0 = \sum_{\alpha\beta}\int dr dr' \bm \phi_{\alpha}^{\dag}(\bm{r})t_{\alpha,\beta}(\bm{r}\!-\!\bm{r}'){\bm \phi}_{\beta}(\bm{r}'),
\end{eqnarray}
%%%%%%%%%%%
where $\alpha$ is the label of the atomic sites $\alpha=d_1$, $d_2$, $p_1$, or $p_2$ and 
${\bm \phi}_\alpha(\bm{r})$ is a spinor composed of the operator ${\phi}_{\alpha}^{s}(\bm{r})$ annihilating 
a state $|\phi_{\alpha}^{s}(\bm{r})\rangle = \sum_{\bm{R}}[\delta(\bm{r}\!-\!\tilde{\bm r}_{\alpha}\!-\!\bm{R})]^{1/2} \left|\alpha, {s} \right>$ 
localized at each atomic site at $\bm{r}=\tilde{\bm{r}}_{\alpha}+\bm{R}$. 
Here, $\bm{R}$ is the lattice translation vector and $|\alpha, s \rangle$ is the wave function of the atomic orbitals 
given in Eqs.~(\ref{eq:d}) and (\ref{eq:p}).
The $t_{\alpha\beta}=[t_{\alpha\beta}]^{s,s'}$ in Eq.~(\ref{eq:tb}) is the $2\times 2$ matrix acting on the spin space.

The spatial symmetry and time-reversal symmetry restrict the hopping parameter. 
Iron chalcogenides have space group symmetry $P4/nmm$, 
whose generators are operators of the point group $C_{4v}$ with the main axis at the chalcogen site
and of inversion with respect to the center of the plaquette (indicated by point $P$ in Fig.~\ref{fig:lattice}). 
The matrix representation of the generator $G$ in the basis $\left|\alpha, s\right>$ is given as 
$[G_{\alpha'\alpha}]^{s's}=\left<\alpha,s \right| G|\alpha's'\rangle$.
Accordingly, we can describe each generator as
%%%%%%%%%%%
\begin{eqnarray} \label{eq:sym}
	\begin{split}
	&C_{4} = e^{- i\frac{3\pi}{4} s_3}\tfrac{\sigma_0+\sigma_3}{2}\eta_1   + e^{-i\frac{\pi}{4} s_3}\tfrac{\sigma_0-\sigma_3}{2}\eta_0  &\\
	&M_{\bm{y}} = -i s_2 \sigma_0 \eta_0 &\\
	&M_{\bm{x}+\bm{y}} =  -i (\tfrac{s_1-s_2}{\sqrt{2}}\tfrac{\sigma_0+\sigma_3}{2}\eta_1 + \tfrac{s_1+s_2}{\sqrt{2}}\tfrac{\sigma_0-\sigma_3}{2}\eta_0) \\
	&P = s_0 \sigma_3 \eta_1.
	\end{split}
\end{eqnarray}
%%%%%%%%%%%
Here, $C_4$ is fourfold rotation, $M_{\bm{n}}$ is a mirror operator with respect to the $\bm{n}=0$ plane, and $P$ is inversion. 
$s_i$, $\sigma_i$, and $\eta_i$ are the $2\times 2$ Pauli matrices acting on the spin, atomic species, 
and sublattice basis, respectively.
Note that $\left|\alpha,s\right>$ with $\alpha=d_{1,2}$ ($p_{1,2}$) is the eigenstate of $\sigma_3$ with the eigenvalue $\lambda_{\sigma_3}=+1$ ($-1$) 
and that with $\alpha=d_{1}$ and $p_1$ ($d_{2}$ and $p_2$) is the eigenstate of $\eta_3$ with $\lambda_{\eta_3}=+1$ ($-1$). 
In other words, we characterize each atomic site $\alpha$ as $(\lambda_{\sigma_3},\lambda_{\eta_3})=(\pm 1,\pm 1)$.
Accordingly, the time-reversal operator is given as 
%%%%%%%%%%%
\begin{eqnarray}\label{eq:T} 
	\mathcal{T} = U_{\mathcal{T}} \mathcal{K} \hbox{ with } U_{\mathcal{T}}= is_2  \sigma_3 \eta_0,  
\end{eqnarray} 
%%%%%%%%%%%
where $\mathcal{K}$ is a complex conjugate operator.

By using the representation~(\ref{eq:sym}) and (\ref{eq:T}), we restrict the hopping parameter to $G_{\alpha\alpha'} t_{\alpha'\beta'}(\bm{r}) G_{\beta'\beta}^\dag=t_{\alpha\beta}(D_{G}[\bm{r}])$, 
where $D_{G}[\bm{r}]$ is the symmetry operation $G$ upon the vector $\bm{r}$.
This symmetry generates all the equivalent nearest and next-nearest neighbor hopping from individual ones depicted in Fig.~\ref{fig:lattice}.
It also restricts the individual hopping as follows:
	$t_{d_2d_1}(a\hat{\bm x}\!+\!a\hat{\bm y}) = t_d s_0$,
	$t_{p_2p_1}(a\hat{\bm x}\!+\!a\hat{\bm y}\!+\!2b \hat{\bm z}) = t_p s_0$,
	$t_{d_1d_1}(0) =t_{d_2d_2}(0)= -t_{p_1,p_1}(0)=-t_{p_2,p_2}(0) = -\delta \mu s_0$, 
	$t_{d_1d_1}(c\hat{\bm z}) =t_{d_2d_2}(c\hat{\bm z})= t_d' s_0$, $t_{p_1,p_1}(c\hat{\bm z}) =t_{p_2p_2}(c\hat{\bm z})= t_p' s_0$,
	and $t_{p_2d_1}(a\hat{\bm x}\!-\![b\!+\!nc] \hat{\bm z}) \equiv t^{n}_{p_2d_1} = t_1^{n} s_2+it_2^{n} s_0$, 
with $n=\pm1$ or 0. The $t_d$, $t_p$, $t_d'$, $t_p'$, $t_1^n$, $t_2^n$ and $\delta\mu$ are real parameters.

%%%%%%%%%%%
\begin{figure}[t]
\begin{center}
	\includegraphics[width=85mm]{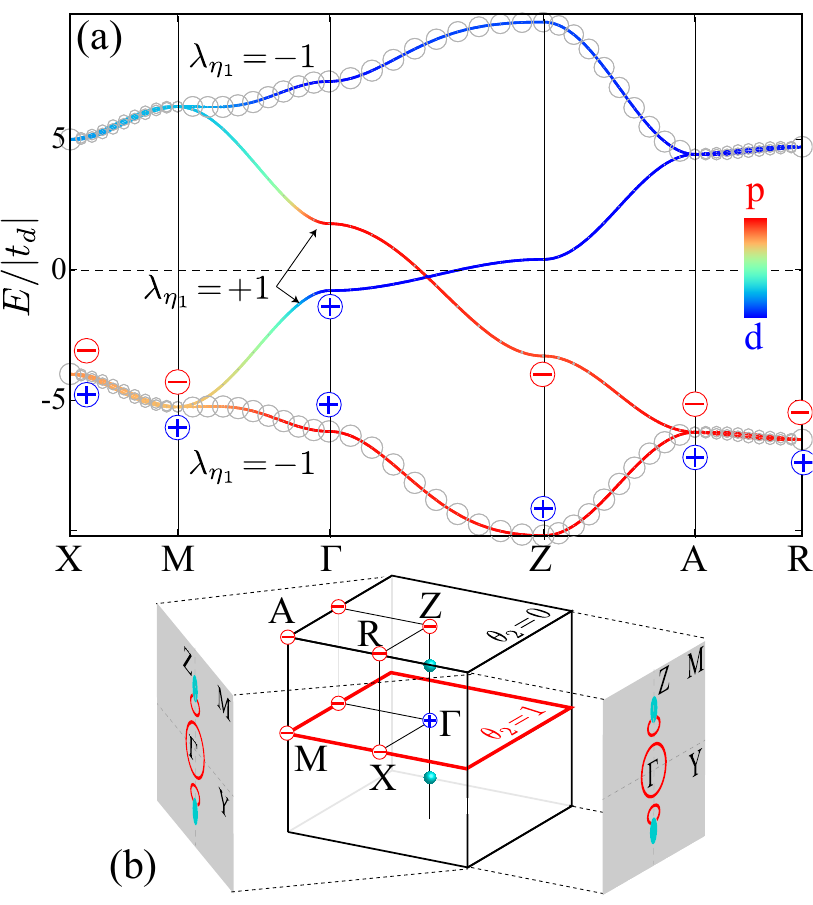}
	\caption{
	(a) Band dispersion obtained from the tight-binding model (\ref{eq:tb}) along a typical high symmetric cut in the Brillouin zone. 
	The color code indicates the orbital property of the band (blue, $d$-orbital; red, $p_z$-orbital; green, mixed).
	The size of the gray circle indicates the weight of the $\lambda_{\eta_1}=-1$ component (see also main text).
	The sign on the band indicates inversion eigenvalue.
	(b) Bulk and surface Brillouin zones with the parity product of the two lower bands at the time-reversal symmetric momenta.  
	The green sphere represents bulk Dirac points. The green shades on the surface Brillouin zone represent the projected Fermi surface. 
	The red curves on the surface Brillouin zone represent the surface Fermi loop.
	The parameters are $t_d=-1.5$ $t_{p}=1$, $t_{d}'=-0.6$, $t_{p}'=1$, $t_1^0=0.5$, $t_2^0=1$, $t_1^\pm = t_2^\pm=0.4\mp 0.2$,   
	$\delta\mu = 2(t_d-t_p)$, and $\mu=-0.5$. 
	}\label{fig:band}
\end{center}
\end{figure}
%%%%%%%%%%%

Foulier transformation of Eq.~(\ref{eq:tb}) gives 
\begin{eqnarray}\label{eq:foulier}
	H_0=\int d\bm{k} {\bm{c}}_{\alpha \bm{k}}^{\dag} H_{0,\alpha\beta}(\bm{k}) {\bm{c}}_{\beta \bm{k}}, 
\end{eqnarray}
where ${\bm{c}}_{\alpha \bm{k}}=\int d\bm{r}e^{i\bm{k}\cdot\bm{r}-ik_z\tilde{z}_{\alpha}} {\bm{\phi}_{\alpha}}(\bm{r})$.  
Diagonalizing $H_{0, \alpha\beta}(\bm{k})$, we obtain the band structure shown in Fig.~\ref{fig:band}.
First, let us focus on the $\Gamma$ and $Z$ points.
These points are invariant under the inversion $P$ and $C_4$ rotation in Eq.~(\ref{eq:sym}), 
and hence, the basis diagonalizing $\eta_1$ is more convenient 
than the sublattice basis diagonalizing $\eta_3$.
The eigenstates with $\lambda_{\eta_1}=\pm 1$ are bonding or antibonding molecular orbitals of sublattice states
$|d_{1}\rangle\pm|d_{2}\rangle$ (also $|p_{1}\rangle\pm|p_{2}\rangle$).
As shown in Table~\ref{table:orbital}, at the $\Gamma$ and $Z$ points, the states with $(\lambda_{\sigma_3} \lambda_{\eta_1})=(\pm 1, \pm 1)$ are
classified in terms of eigenvalues $\lambda_{P}$ and $\lambda_{C_4}=e^{i m_4 \frac{\pi}{2}}$.
Each state corresponds to an irreducible representation of the space group $P4/nmm$ at the $\Gamma$ and $Z$ points.

In Fe(Se,Te), band inversion between the $d$-orbital with $(\lambda_{\sigma_3},\lambda_{\eta_1})=(+1,+1)$ at the iron atom and the $p_z$-orbital with $(-1,+1)$ 
occurs on the $\Gamma Z$ path. 
These states are characterized by different $C_4$ eigenvalues with $|m_4|= 3/2$ and $1/2$. 
As momenta along the $\Gamma Z$ path are invariant under the $C_4$ rotation, these states cannot hybridize with each other. 
Therefore, the energy-crossing point on the path remains a Dirac point. 

It is also worth comparing the energy of the molecular orbitals $\lambda_{\eta_1}=-1$ with that of $\lambda_{\eta_1}=1$. See Fig.~\ref{fig:band}. 
For the $d$-orbital,  $\lambda_{\eta_1}=-1$ is the antibonding molecular orbital, 
and hence has higher energy than $\lambda_{\eta_1}=+1$.
Meanwhile, for the $p_z$-orbital, $\lambda_{\eta_1}=-1$ is the bonding molecular orbital,
because the two sublattices of the chalcogen sites are displaced from the iron plane oppositely.
Therefore, it has lower energy than $\lambda_{\eta_1}=+1$.
%%%%%%%%%%%
\begin{table}[b]
	\caption{Inversion eigenvalues $\lambda_P$, $C_4$, rotation eigenvalues $\eta_{C_4}=e^{i m_4 \pi/2}$, 
	and irreducible representation of the space group $P4/nmm$ for eight different bands at the $\Gamma$ and $Z$ points.
	The eigenvalues $\lambda_{\sigma_3}=\pm$ represent $d$ and $p$ orbital.}\label{table:orbital}
	\renewcommand{\arraystretch}{1.2}
	\newcolumntype{P}[1]{>{\centering\arraybackslash}p{#1}}
	\begin{tabularx}{85mm}{ P{25mm} cccc}
	\hline \hline
	$(\lambda_{\sigma_3} \lambda_{\eta_1},\lambda_{s_3})$ & $(+,+,\pm)$ & $(+,-,\pm)$ & $(-,+,\pm)$ & $(-,-,\pm)$ \\
	\hline
	$\lambda_{P}$ & $+1$ & $-1$ & $-1$ & $+1$ \\
	$m_4$    & $\pm 3/2$  & $\pm 1/2$  & $\pm 1/2$  & $\pm 1/2$ \\
	\rule{0pt}{3ex}
	$P4/nmm$ & $E_{3/2g}$ & $E_{1/2u}$ & $E_{1/2u}$ & $E_{1/2g}$ \\
	\hline	\hline
	\end{tabularx} 
\end{table}
%%%%%%%%%%%

By using the obtained level structure, we evaluate the topological index of the Dirac point.
We can consider the $k_z=0$ and $k_z=\pi$ plane in momentum space as a two-dimensional system 
with time-reversal and inversion symmetries.
Therefore, the topological invariant for these planes is the $\mathbb{Z}_2$ index introduced by Kane and Mele~\cite{kane2005_Z2}. 
We evaluate it as a parity product of the occupied state at the time-reversal momenta~\cite{Fu2007} as 
\begin{eqnarray}\label{eq:z2}
	e^{i \pi\theta_2(k_z)} = \prod_{E_n<0} \lambda_P^{n} (0,0,k_z)\lambda_P^{n} (\pi,\pi,k_z), 
\end{eqnarray}
where $\lambda_P^{n} (k_x,k_y,k_z)$ is the parity eigenvalue at the time-reversal invariant momentum
also shown in Fig.~\ref{fig:band}.
From the energy dispersion and the parity shown in Fig.~\ref{fig:band}, 
we observe that the $\mathbb{Z}_2$ indices are $\theta_2(0)=1$ and $\theta_2(\pi)=0$. 
As the $k_z=0$ plane is topologically non-trivial, 
the Fermi loop appears around $k_z=0$ of the surface Brillouin zone~\cite{yang2014} as shown in Fig.~\ref{fig:band} (b).

\begin{figure}[t]
\begin{center}
	\includegraphics[width=85mm]{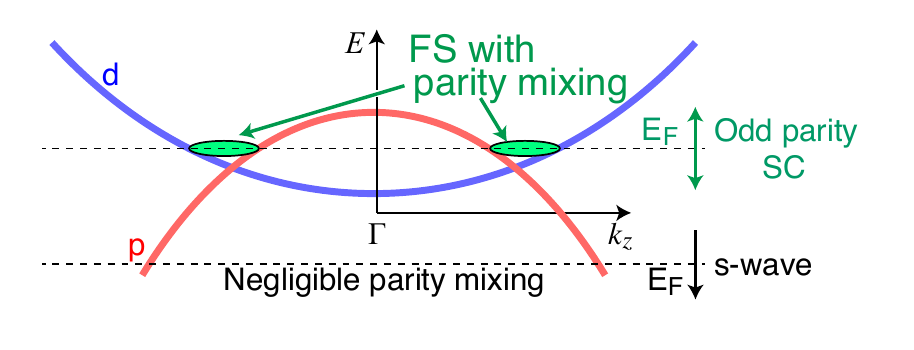}
	\caption{
	Schematics of the band structure and position of the Fermi level. 
	In the case where the Fermi level is close to the Dirac points, odd-parity superconductivity is possible. 
	}\label{fig:phasediag}
\end{center}
\end{figure}

\section{Possible superconducting gap}\label{sec:super}
Let us consider the superconductivity of the obtained Dirac semimetal.
When the Fermi level is far from the energy of the Dirac point (see Fig.~\ref{fig:phasediag}), 
the parity mixing on the Fermi surface is negligibly weak. 
Hence, the only possible superconductivity is the $s$-wave pairing state.
The $s$-wave pair is topologically trivial in the bulk, 
but it may induce a topological superconductor on the surface of system. 
Recently, the presence of Majorana bound states in a vortex of this surface superconductor 
has been theoretically discussed~\cite{gang2017,qin2019,konig2019}. 
By contrast, when the Fermi level is close to the Dirac points, 
opposite-parity states strongly mix with each other on the Fermi surfaces [see Fig.~\ref{fig:phasediag} and~\ref{fig:gap}(a)].
This hybridization naturally allows the Cooper pair 
between electrons with opposite orbital parity. 
Here, we discuss the possible odd-parity pairing induced from this pairing and its topological property.

Here, we use the Bogoliubov--de Gennes (BdG) formalism within the Nambu space $(\bm{c}_{\bm k}, \bar{\bm c}_{-{\bm k}})$. 
The spinor $\bar{\bm{c}}_{-\bm{k}}=U_{\mathcal{T}}\bm{c}_{-\bm{k}}^\dag$ with $U_{\mathcal{T}}$ defined in Eq.~(\ref{eq:T}) is the time-reversal hole partner 
of annihlation operator $\bm{c}_{\bm{k}}$.
Accordingly, the BdG Hamiltonian is written as 
\begin{eqnarray}\label{eq:bdg}
	H(\bm{k}) = 
	\left( \begin{array}{cc}
		H_0(\bm{k}) & \Delta(\bm{k}) \\
		\Delta^\dag(\bm{k}) & -H_0(\bm{k})
	\end{array}\right), 
\end{eqnarray}
where $H_0(\bm{k})$ is a one-particle Hamiltonian given in Eq.~(\ref{eq:foulier}) and we use the time-reversal symmetry $\mathcal{T}H_0(\bm{k})\mathcal{T}^{-1}=H_0(-\bm{k})$.

Subsequently, we consider the possible gap functions.
The present system involves spin, orbital, and sublattice d.o.f. 
Then, we can generally describe the gap function as
\begin{eqnarray}
	\Delta(\bm{k}) = s_{\mu}\sigma_{\nu}\eta_{\gamma} f(\bm{k}).
\end{eqnarray}
The spin component $s_{\mu}$ can take the values $\mu=0,1,2,3$. 
Here, we focus on the pairing between the $d$- and $p_z$-orbitals with $\sigma_\nu=\sigma_1$ or $\sigma_2$.
It is the pairing between the nearest neighboring sites. 
Sublattices 1 and 2 appear alternately along the $x$ direction in the present model, as shown in Fig.~\ref{fig:lattice}. 
Therefore, the (off-)diagonal matrices $\eta_3$ and $\eta_0$ ($\eta_1$ and $\eta_2$) in the sublattice basis indicate 
the pairing between the neighboring sites in the $\bm{y}$ ($\bm{x}$) direction.
This pairing direction restricts the possible form of the momentum dependence $f(\bm{k})$ coupling with $\eta_\nu$ as
$\eta_\gamma f(\bm{k}) = \eta_{1,2} \cos(k_x a), \eta_{1,2} \sin(k_x a), \eta_{3,0} \cos(k_y a), \eta_{3,0} \sin(k_y a) 
\equiv \eta_{1,2}^{\mathrm{c}}, \eta_{1,2}^{\mathrm{s}},\eta_{3,0}^{\mathrm{c}}, \eta_{3,0}^{\mathrm{s}}$. 
We can combine the spin, orbital, and sublattice components 
to satisfy the Fermi statistics $(\Delta(\bm{k})U_{\mathcal{T}})^\mathcal{T}=-\Delta(\bm{k})U_{\mathcal{T}}$. 
Consequently, we obtain the 32 possible pairing functions in Table~\ref{table:gap} 
and classify them in terms of the irreducible representations of the space group $P4/nmm$.

\begin{table}[t]
	\caption{Possible gap function for the nearest-neighbor pairing states, 
	where the abbreviated notations are $\eta_{1,2}^{\mathrm{c}}=\eta_{1,2} \cos k_x a$, $\eta_{1,2}^{\mathrm{s}}=\eta_{1,2} \sin k_x a$, 
	$\eta_{3,0}^{\mathrm{c}}=\eta_{3,0} \cos k_y a$, and $\eta_{3,0}^{\mathrm{s}}=\eta_{3,0} \sin k_y a$.
	} \label{table:gap}
	\renewcommand{\arraystretch}{1.2}
	\newcolumntype{C}{>{\centering\arraybackslash}X}
	\newcolumntype{P}[1]{>{\centering\arraybackslash}p{#1}}
	\begin{tabularx}{85mm}{  P{5mm} P{15mm}|C}
	\hline \hline
	 $C_{4v}$ & $P4/nmm$ &$\Delta(\bm{k})$ \\
	\hline
	  $A_{1}$ & $A_{1g}$ &$s_1(\sigma_2 \eta_2^{\mathrm{c}} \!-\! \sigma_1 \eta_3^{\mathrm{c}})$,\quad $s_3 \sigma_2\eta_1^{\mathrm{s}} \!-\! s_0 \sigma_1\eta_0^{\mathrm{s}}$ \\
	          & $A_{2u}$ &$s_1 \sigma_1 (\eta_{1}^{\mathrm{c}}  \!-\! \eta_{0}^{\mathrm{c}})$,\quad $(\eta_{2}^{\mathrm{s}}s_3 \!+\! \eta_{3}^{\mathrm{s}}s_0)\sigma_1$ \\
	  \hline
	  $A_{2}$ & $A_{2g}$ &$s_2(\sigma_2 \eta_2^{\mathrm{c}} \!-\! \sigma_1 \eta_3^{\mathrm{c}})$,\quad $s_0\sigma_1\eta_1^{\mathrm{s}} \!+\!  s_3\sigma_2\eta_0^{\mathrm{s}}$ \\
	          & $A_{1u}$ &$ s_2 \sigma_1 (\eta_1^{\mathrm{c}}  \!-\!\eta_0^{\mathrm{c}})$,\quad $(\eta_2^{\mathrm{s}}s_0  \!+\! \eta_3^{\mathrm{s}}s_3)\sigma_2 $ \\
	  \hline
	  $B_{1}$ & $B_{1g}$ &$s_1(\sigma_2 \eta_2^{\mathrm{c}} \!+\! \sigma_1 \eta_3^{\mathrm{c}})$,\quad $s_3 \sigma_2\eta_1^{\mathrm{s}} \!+\! s_0 \sigma_1\eta_0^{\mathrm{s}}$ \\
	          & $B_{2u}$ &$s_2 \sigma_1 (\eta_1^{\mathrm{c}}  \!+\!\eta_0^{\mathrm{c}})$,\quad $(\eta_2^{\mathrm{s}}s_0  \!-\! \eta_3^{\mathrm{s}}s_3)\sigma_2 $ \\
	  \hline
	  $B_{2}$ & $B_{2g}$ &$s_2(\sigma_2 \eta_2^{\mathrm{c}} \!+\! \sigma_1 \eta_3^{\mathrm{c}})$,\quad $s_0\sigma_1\eta_1^{\mathrm{s}} \!-\!  s_3\sigma_2\eta_0^{\mathrm{s}}$ \\
	          & $B_{1u}$ &$s_1 \sigma_1 (\eta_1^{\mathrm{c}}  \!+\!\eta_0^{\mathrm{c}})$,\quad $(\eta_{2}^{\mathrm{s}}s_3 \!-\! \eta_{3}^{\mathrm{s}}s_0)\sigma_1$ \\
	  \hline
	  $E$     & $E_g$    &$\sigma_1\{  \eta_2^{\mathrm{c}}s_0,\  \eta_3^{\mathrm{c}}s_3  \},\quad \sigma_2\{  \eta_3^{\mathrm{c}}s_0,\  \eta_2^{\mathrm{c}}s_3  \}$ \\
	          &          &$s_1 \sigma_2 \{  \eta_0^{\mathrm{s}},\  \eta_1^{\mathrm{s}} \},\quad \ s_2 \sigma_2 \{  \eta_0^{\mathrm{s}},\  \eta_1^{\mathrm{s}} \}$\\
	          & $E_u$    &$s_1\{  \sigma_1 \eta_2^{\mathrm{s}},\ \sigma_2 \eta_3^{\mathrm{s}} \},\quad s_2\{  \sigma_1 \eta_2^{\mathrm{s}},\ \sigma_2 \eta_3^{\mathrm{s}} \}$ \\
	          &          &$\{ s_3\sigma_1 , \ s_0 \sigma_2 \}({\eta_1^{\mathrm{c}}\pm\eta_0^{\mathrm{c}}})$ \\
		\hline
		\hline
	\end{tabularx} 
\end{table}

Let us choose the gap functions with a higher transition temperature $T_c$,
which correspond to those with relatively larger superconducting gaps on the Fermi surface at zero temperature.
First, for the small Fermi surface around the Dirac point at $k_x=k_y=0$, which is considered here, 
the gap function with the sinusoidal $k$-dependence $\eta_{\gamma}\sin (k_{i}a)$ $(i=x,y)$ is rather small.
Subsequently, we check the sublattice d.o.f. 
in terms of the molecular orbital basis diagonalizing $\eta_1$. 
Accordingly, the $\Delta\propto \eta_1\!+\!\eta_0=\mathrm{diag(1,0)}$ 
($\Delta\propto \eta_1\!-\!\eta_0=\mathrm{diag(0,1)}$) opens a gap on the Fermi surface of the $\lambda_{\eta_1}=+1$ ($-1$) bands.
Those with $\Delta\propto \eta_{2,3}$, off-diagonal on this basis, can open the gap when the two states with different 
$\lambda_{\eta_1}$ mix with each other on the Fermi surface.   
In the present system, the bands near the Dirac point contain almost no $\eta_1=-1$ component (See Fig.~\ref{fig:band}). 
Therefore, the gap function with the sublattice component $\eta_1 + \eta_0$ opens an energy gap larger than the others. 
From the above arguments, we narrow down the candidate of irreducible representation with higher $T_c$ to $B_{1u}$, $B_{2u}$, or $E_{u}$ in Table~\ref{table:gap}. 
Note that these representations coincide with those of odd-parity superconductivity obtained in the $k\cdot P$ model of the Dirac semimetal accompanied with contact pairing interaction~\cite{hashimoto2016}.

\begin{figure}[b]
\begin{center}
	\includegraphics[width=85mm]{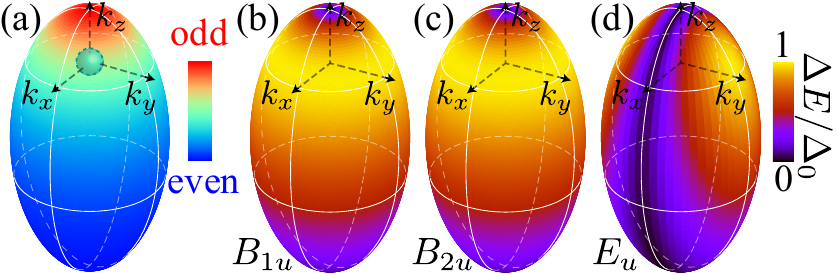}
	\caption{
	(a) Fermi surface around the Dirac point, where the color code indicates the expectation value of the parity operator $P$ on it.
	(b-d) The normalized superconducting gap on the Fermi surface of the $B_{1u}$, $B_{2u}$, and $E_u$ representations.
	}\label{fig:gap}
\end{center}
\end{figure}

Figure~\ref{fig:gap} shows the energy gap on the Fermi surface. 
For all the three gap functions, we have gap nodes at two poles on the $k_z$ axis. 
The energy gap of the $E_{u}$ representation is also suppressed along the lines connecting them. 
Therefore, the $B_{1u}$ and $B_{2u}$ representations support higher $T_c$ than the $E_{u}$. 
The $B_{1u}$ and $B_{2u}$ representations differ only in the
character of the diagonal and vertical mirror reflections; 
for instance, the gap function satisfies the commutation or anticommutation relation
$[M_{\bm{y}}, \Delta]_{\mp}=0$ and $[M_{\bm{x}\!+\!\bm{y}}, \Delta]_{\pm}=0$ where the upper (lower) sign is for $B_{1u}$ ($B_{2u}$). 
However, the in-plane anisotropy between the vertical (100) and diagonal (110) directions is small. 
Therefore, $T_c$ for these two representations is almost degenerate. 

\section{Topological properties of the $B_{1u}$ and $B_{2u}$ states }
We examine the symmetry-protected topological properties of the possible superconducting states $B_{1u}$ and $B_{2u}$.
The BdG Hamiltonian has particle-hole symmetry,
\begin{eqnarray}\label{eq:PHS}
	{\mathcal{C}}H(\bm{k}){\mathcal{C}}^{-1} =-H(-\bm{k}) \hbox{ with } \mathcal{C}= \tau_2 U_{\mathcal{T}}\mathcal{K},
\end{eqnarray}
inherent to superconductors. Here, $\tau_{\nu=0,1,2,3}$ represents the identity and Pauli matrices acting on the particle and time-reversal hole space 
and $U_{\mathcal{T}}$ is the unitary part of the time-reversal operator [see Eq.(\ref{eq:T})]. 
In addition, both the $B_{1u}$ and $B_{2u}$ states preserve the time-reversal symmetry 
\begin{eqnarray}\label{eq:Ts}
	\tilde{\mathcal{T}}H(\bm{k})\tilde{\mathcal{T}}^{-1} = H(-\bm{k}) \hbox{ with } \tilde{\mathcal{T}}= \tau_0 U_{\mathcal{T}}\mathcal{K}.
\end{eqnarray}
By combining them, we obtain chiral symmetry 
\begin{eqnarray}\label{eq:chiral}
	\gamma H(\bm{k})\gamma^{-1} = -H(\bm{k}) \hbox{ with } \gamma= \tilde{\mathcal{T}}\mathcal{C} = \tau_2.
\end{eqnarray}
We also consider the crystalline symmetry of the BdG Hamiltonian, formally given as
\begin{eqnarray}\label{eq:Gsc}
	\tilde{G}H(\bm{k})\tilde{G}^{-1} = H(D_{G}[\bm{k}]),
\end{eqnarray}
for the generator $G$ of the space group symmetry. 
Note that the unitary operator $\tilde{G}$ acting on the Nambu space and its commutation or anticommutation relation with the operators 
in Eq.~(\ref{eq:PHS}), (\ref{eq:Ts}), and (\ref{eq:chiral}) depend on the  
irreducible representations of the gap function (see later discussion). 
We can define rich topological numbers by using these relations.
Owing to these topological numbers, the system supports the bulk--surface correspondence
depending on the irreducible representations of the gap function and the direction of the surface.

\begin{figure}[b]
\begin{center}
	\includegraphics[width=85mm]{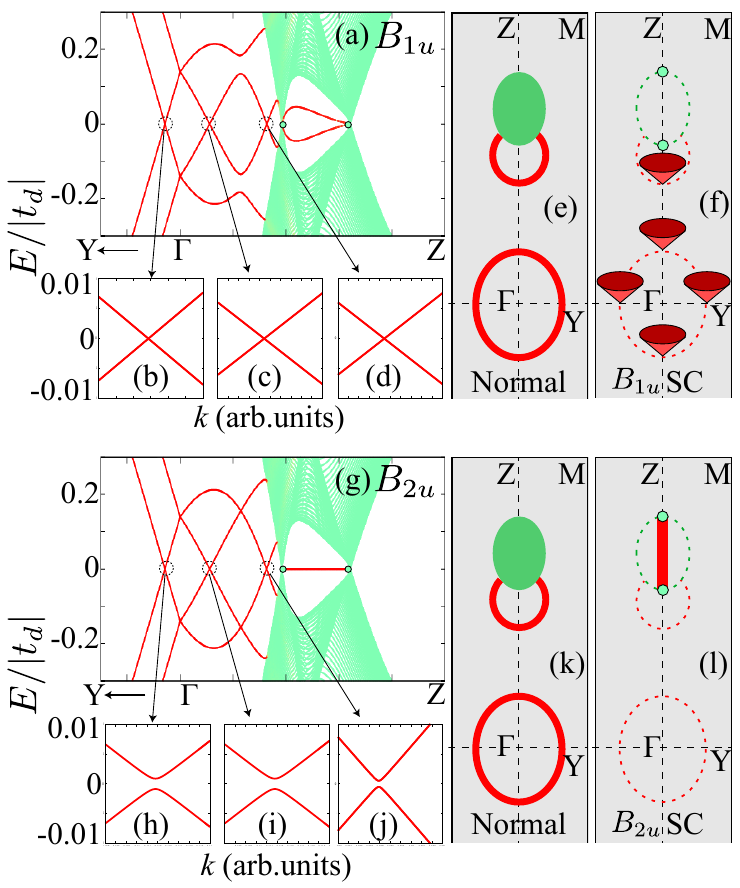}
	\caption{
	Energy spectrum of the system with the $(100)$ surface. 
	(a-f) indicate the $B_{1u}$ states. (a) shows the dispersion along a typical high symmetric cut in the surface Brillouin zone.
	The bands in green and red indicate the bulk and surface states, respectively. 
	(b-d) magnify the momentum regions around the surface Fermi loop.
	(e) shows the schematics of the surface zero-energy states on the surface Brillouin zone in the normal state. 
	The projected bulk Fermi surface and the surface Fermi loop are depicted by green shade and red curve, respectively.
	(f) is the the surface zero energy states in the $B_{1u}$ superconducting state. 
	In (f), the bulk Fermi surface and Fermi loop in the normal state  (indicated by green and red dashed curves, respectively) are gapped at generic point. 
	At intersections with mirror invariant lines, the surface Fermi loop survives as Majorana fermions in the superconducting state.
	Around the Gamma point, they form a Majorana quartet~\cite{kobayashi2015}
	(g-l) is the same plot as (a-f) but for the $B_{2u}$ state. The thick red line in (l) shows the Majorana flat band. 
	The parameters are the same as Fig.~\ref{fig:band}. 
	}\label{fig:100surf}
\end{center}
\end{figure}

\begin{figure}[b]
\begin{center}
	\includegraphics[width=85mm]{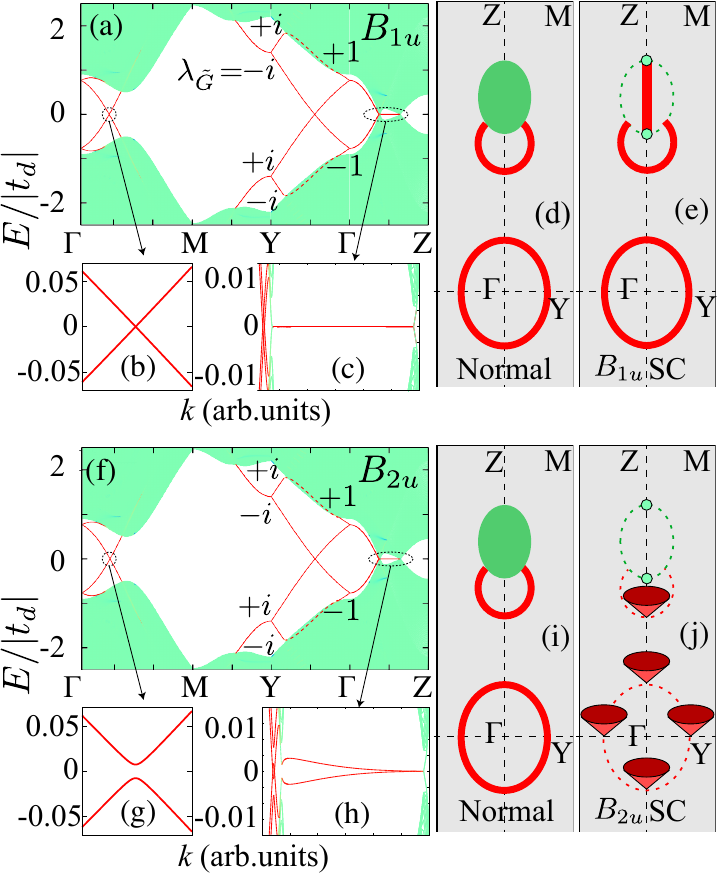}
	\caption{
	Energy spectrum of the system with the $(110)$ surface in a manner similar to Fig.~\ref{fig:100surf}. 
	(a-e) and (f-j) indicate the $B_{1u}$ and $B_{2u}$ states, respectively. 
	Contrary to $(100)$ surface shown in Fig.~\ref{fig:100surf}, 
	the $B_{1u}$ state exhibits a flat band on the $\Gamma Z$ path 
	and the $B_{2u}$ has a Majorana quartet.
	In addition, all the surface Fermi loop remains gapless for the $B_{1u}$ superconductor, 
	associated with the diagonal $C_2''$-odd property of the gap function.
	Both of the $B_{1u}$ and $B_{2u}$ state have M{\"o}bius twisted surface states (hourglass fermions),  
	passing through the bulk spectrum as shown in dashed line in (a) and (f). 
	They are typical for glide protected time-reversal invariant topological phases. (see main text for more details)
	The parameters are the same as  Figs.~\ref{fig:band} and ~\ref{fig:100surf}.
	}\label{fig:110surf}
\end{center}
\end{figure}

To exhaust these topological properties, 
we analyze the surface spectrum of the $B_{1u}$ and $B_{2u}$ states by numerically solving the BdG equation.
We summarize the observed gapless states on the (100) and (110) surfaces in Fig.~\ref{fig:100surf} and \ref{fig:110surf},
and we will discuss their characteristics and topological origin below.

\subsection{Majorana Flat Bands}
We observe the flat zero-energy bands along the $\Gamma Z$ path of the (100) surface of $B_{2u}$ [Fig.~\ref{fig:100surf} (g) and (l)] 
and (110) of $B_{1u}$ [Fig.~\ref{fig:110surf} (c) and (e)]. 
They originate from bulk one-dimensional (1D) topological numbers.
Specifically, let us consider those in the $B_{2u}$ states shown in Fig.~\ref{fig:100surf}(f). 
The $B_{2u}$ gap function is symmetric under the vertical mirror reflections $M_{\bm y}\Delta(\bm{k})M_{\bm y}^{-1}=\Delta(D_{M}[\bm{k}])$. 
Hence, the mirror reflection operator acting on the Nambu space is given as
\begin{eqnarray}\label{eq:Mb2u}
	\tilde{M}_{\bm{y}}=M_{\bm y}\tau_0.
\end{eqnarray}
From Eq.(\ref{eq:Gsc}) with $G=M_{\bm{y}}$ and $D_{M_{\bm y}}[k_x,k_y,k_z]=(k_x,-k_y,k_z)$, 
we obtain the commutation relation $[H(k_x,0,k_z), \tilde{M}_{\bm y}]=0$. 
Therefore, in basis diagonalizing the mirror operator as $U_{\tilde{M}_{\bm{y}}}\tilde{M}_{\bm_{y}}U_{\tilde{M}_{\bm_{y}}}^{-1}=\mathrm{diag}(+i,-i)$, 
the Hamiltonian is also diagonal as $U_{\tilde{M}_{\bm_{y}}}H(k_x,0,k_z)U_{\tilde{M}_{\bm_{y}}}^{-1}=\mathrm{diag}(H_{+i},H_{-i})$. 

The mirror operator in Eq.~(\ref{eq:Mb2u}) also commutes with the chiral operator in Eq.~(\ref{eq:chiral}). 
Hence, $U_{\tilde{M}_{\bm_{y}}}{\gamma}U_{\tilde{M}_{\bm_{y}}}^{-1}=\mathrm{diag}(\gamma_{+i},\gamma_{-i})$. 
The simultaneous diagonalizability of $H(k_x,0,k_z)$, $\tilde{M}_{\bm{y}}$, and $\gamma$ 
indicates that each block of the Hamiltonian has chiral symmetry
\begin{eqnarray}\label{eq:chM}
  \gamma_{\pm i} H_{\pm i}(k_x,0,k_z) \gamma_{\pm i}^{-1} = -H_{\pm i}(k_x,0,k_z). 
\end{eqnarray}
By using this symmetry, we can introduce the 1D winding number
\begin{eqnarray}\label{eq:w1d}
	w_{\pm i}(k_z)= -\frac{1}{4\pi i}\int d{k_x} \mathrm{tr}[\gamma^{\pm i} (H_{\pm i}(k_x,0,k_z))^{-1} \nn\\(\partial_{k_{x}}H_{\pm i}(k_x,0,k_z))]. 
\end{eqnarray}
Note that the total winding number is always zero $w=w_{+i}+w_{-i}=0$ 
for the odd-parity superconductivity~\cite{kobayashi2014}, 
but the mirror winding number $w_M=(w_{+i}-w_{-i})/2$ can be non-trivial.
In the Dirac semimetallic state shown in Fig.~\ref{fig:band} coupled with the $B_{2u}$ gap function, 
we numerically determine that 
\begin{eqnarray}\label{eq:w1dval}
	w_{M}(k_z)= 
	\left\{ \begin{array}{l}
	1 \hbox{ for } k_1 < k_z < k_2 \\
	0 \hbox{ otherwise }
	\end{array}\right.,
\end{eqnarray}
where $k_{i=1,2}$ represents the momenta of the north and south poles of the Fermi surface $\bm{k}=\pm k_i \hat{\bm z}$.
The non-trivial mirror winding number in Eq.~(\ref{eq:w1dval}) ensures the existence of the zero-energy mode on 
$k_y=0$ and between the projected point nodes on the surface perpendicular to the $x$-axis, 
consistent with Fig.~\ref{fig:100surf}(g) and (l).

Similarly, $B_{1u}$ is symmetric under the diagonal mirror reflection as $M_{-\bm{x}+\bm{y}}\Delta(\bm{k})M_{-\bm{x}+\bm{y}}^{-1}=\Delta(D_{M_{-\bm{x}+\bm{y}}}[\bm{k}])$. 
Thus, the above discussion is directly applicable to $B_{1u}$ by replacing $x \rightarrow x+y$ , $y\rightarrow -x+y$. 
Hence, the $B_{1u}$ state has a zero-energy flat band on the diagonal $(110)$ surface as shown in Fig.~\ref{fig:110surf} (c) and (e). 

By contrast, the $B_{2u}$ ($B_{1u}$) gap function is odd under diagonal (vertical) mirror reflection. 
In this case, we cannot define the chiral symmetry (\ref{eq:chM}) in a mirror sector and the mirror winding number (\ref{eq:w1d}),  
and hence, the surface states between the projected point nodes on the (110) [(100)] surface of the $B_{2u}$ ($B_{1u}$) states split 
as shown in Figs.~\ref{fig:100surf}(a) and ~\ref{fig:110surf}(h). 

\subsection{Point and Line Nodes on Surface Fermi Loop}
In addition to the Majorana flat band ensured by the bulk 1D winding number~(\ref{eq:w1d}), 
we observe other gapless spectra on the surface Fermi loop (see Fig.~\ref{fig:100surf} and \ref{fig:110surf}).
We reveal in this section that the topological numbers defined by the surface state characterize them.
Let us start our discussion with the (100) surface of the $B_{1u}$ state. 
As shown in Fig.~\ref{fig:100surf}(a-f), the surface state passes through the
zero-energy points on the $\Gamma Y$ and $\Gamma Z$ paths.
The key to understanding them is vertical mirror reflection symmetry.

First, we examine the gapless states on the $\Gamma Y$ path. 
The effective Hamiltonian of the (100) surface state is written as $H_{100}(k_y,k_z)=H_{0,100}({k_y,k_z})\tau_3 + \Delta_{100}(k_y,k_z)\tau_1$. 
Here, $H_{0,100}(k_y,k_z)$ is the surface Hamiltonian of the normal state and $\Delta_{100}(k_y,k_z)$ is 
the gap function projected on the surface state. 
As the bulk $B_{1u}$ gap function is odd under the vertical mirror reflection $M_{\bm y}$, that on the surface also satisfies 
$M_{\bm y}\Delta_{100}(k_y,k_z)M_{\bm y}^{-1}=-\Delta_{100}(D_{M_{\bm{y}}}[k_y,k_z])$. 
In this case, the mirror symmetry operator acting on the Nambu space is given as 
\begin{eqnarray}\label{eq:Modd}
	\tilde{M}_{\bm{y}} = M_{\bm{y}}\tau_{3}.
\end{eqnarray}
The combination of the mirror reflection symmetry (\ref{eq:Modd}) and particle-hole symmetry (\ref{eq:PHS}) yields 
the antiunitary antisymmetry of the BdG Hamiltonian, 
\begin{eqnarray}\label{eq:phM}
	\mathcal{C}_{\tilde{M}_{\bm{y}}} H_{100}(k_y,k_z) \mathcal{C}_{\tilde{M}_{\bm{y}}}^{-1} = -H_{100}(k_y,-k_z).
\end{eqnarray}
with 
\begin{eqnarray}\label{eq:CM}
\mathcal{C}_{\tilde{M}_{\bm y}} = \tilde{M}_{\bm y}\mathcal{C} = -\tau_1 M_{\bm y} U_{\mathcal{T}}\mathcal{K} .
\end{eqnarray}
The operator in Eq.~(\ref{eq:Modd}) of the mirror-odd superconductor satisfies the anticommutation relation $\{\tilde{M}_{\bm y}, \mathcal{C}\}=0$. 
This relation together with $\mathcal{C}^2=1$, $\tilde{M}_{\bm{y}}^2=-1$ yields $\mathcal{C}^2_{\tilde{M}_{\bm y}}=1$. 
That is, we can consider the surface state with fixed $\bm{k}$ on $k_z=0$ as a zero-dimensional system 
in the class $D$ with particle-hole symmetry $\mathcal{C}_{\tilde{M}_{\bm y}}$.
In this case, Eq.~(\ref{eq:phM}) and (\ref{eq:CM}) indicate that $H_{100}(k_y,0)i\tau_1 M_{\bm{y}}U_{\mathcal{T}}$ is unitary equivalent to a real and antisymmetric matrix.
Thus, we can immediately introduce the $\mathbb{Z}_2$ topological number 
\begin{eqnarray}
	\chi(k_y)=\mathrm{sgn}\{ \mathrm{Pf}[H_{100} i\tau_1 M_{\bm y} U_{\mathcal{T}}] \}. 
\end{eqnarray}
In the weak coupling limit $\Delta(k_y,k_z)\ll E_{\mathrm{F}}$, 
we can evaluate this number from a particle Hamiltonian as $\chi(k_y)=\mathrm{sgn}[\mathrm{det}(H_{0,100})]$.
Thus, the Fermi loop, where the sign of an eigenvalue of $H_{0,100}$ changes,
is the boundary between the regions with opposite-signed $\chi(k_y)$. 
Owing to the difference of $\chi(k_y)$, the zero-energy modes on the $\Gamma Y$ path in Fig.~\ref{fig:100surf}(b) and (f) appear.

Subsequently, let us show that symmetry of the surface Hamiltonian also protects 
the zero-energy states on the $\Gamma Z$ path in Fig.~\ref{fig:100surf} (c), (d), and (f). 
We start with a non-interacting case with $\Delta=0$, where the particle ($\lambda_{\tau_3}=1$) 
and time-reversal hole part ($\lambda_{\tau_3}=-1$) are completely decoupled.
In this case, a particle state $\tilde{\bm u}_{n,\bm{k}}=(\bm{u}_{n,\bm{k}}^{T},\bm{0}^T)^T$ with eigenenergy $E_{n}$ can be a solution of the BdG equation.
By contrast, the chiral symmetry~(\ref{eq:chiral}) ensures that a hole state 
$\tilde{\bm u}_{n',\bm{k}}=\gamma \tilde{\bm u}_{n,\bm{k}}=(\bm{0}^T, i\bm{u}_{n,\bm{k}}^{T})^T$ with eigenenergy $E_{n'}=-E_{n}$
is also a solution. 
While the energy dispersion of $\tilde{\bm u}_{n,\bm{k}}$ and $\gamma\tilde{\bm u}_{n,\bm{k}}$ may cross at Fermi level $E_n=0$ when $\Delta=0$, 
weak, but finite, coupling $\Delta\ll E_F$ opens an energy gap at the crossing point at the generic momentum $\bm{k}$.

However, the mirror symmetry~(\ref{eq:Modd}) of this system prohibits the energy gap on the $\Gamma Z$ path. 
As this path is invariant under vertical mirror reflection,
the eigenstates of the BdG Hamiltonian are also those of the mirror operator simultaneously, 
\begin{eqnarray}
\tilde{M}_{\bm{y}}\tilde{\bm u}_{n,k_z} = \lambda_{\tilde{M}_{\bm y}}^{(n)} \tilde{\bm u}_{n,k_z},
\end{eqnarray}
where $\tilde{\bm u}_{n,k_z}$ is the wave function $\tilde{\bm u}_{n,\bm{k}}$ on $k_y=0$.
Owing to the anticommutation relation $\{\gamma, \tilde{M}_{\bm y}\}=0$ of mirror-odd superconductivity,
the particle and time-reversal hole solutions $\tilde{\bm u}_{n,k_z}$ and $\tilde{\bm u}_{n',k_z}=\gamma \tilde{\bm u}_{n,k_z}$ 
have different mirror eigenvalues $\lambda^{(\nu)}_{\tilde{M}_{\bm y}}=-\lambda^{(\nu')}_{\tilde{M}_{\bm y}}$. 
Therefore, on the $\Gamma Z$ path, these two states do not interact with each other even when $\Delta$ is finite, 
and energy crossing at zero energy remains as shown in Fig.~\ref{fig:100surf} (c), (d), and (f).

In contrast to the $B_{1u}$ state, the $B_{2u}$ state is odd under the diagonal mirror reflection ${M}_{-\bm{x}+\bm{y}}$. 
Applying the above discussion to $B_{2u}$ states by replacing $\bm{y}$ with $-\bm{x}+\bm{y}$ and $H_{100}$ with $H_{110}$,
we can conclude that the zero-dimensional topological number of the (110) surface states characterizes
the zero-energy points on the Fermi loop in Fig.~\ref{fig:110surf}(j). 

In addition, as shown in Fig.~\ref{fig:110surf}(a)-(e), 
the energy gap closes everywhere on the surface Fermi loop for the $B_{1u}$ state with the (110) surface. 
It also originates from the symmetry-protected topological number of the surface states.
The relevant symmetry is twofold rotation symmetry $C''_2$ about a diagonal axis as depicted in Fig.~\ref{fig:lattice}. 
As the $B_{1u}$ state is odd under this $C''_2$ rotation, 
we have the antiunitary antisymmetry of the BdG Hamiltonian
\begin{eqnarray}
	\mathcal{C}_{\tilde{C}_{2}''} H_{110}(k_{\bar{1}10},k_z) \mathcal{C}_{\tilde{C}_{2}''}^{-1} = -H_{110}(k_{\bar{1}10},k_z), 
\end{eqnarray}
with $\mathcal{C}_{\tilde{C}_{2}''} = i\tilde{C}_{2}''\mathcal{C}$, $\mathcal{C}_{\tilde{C}_{2}''}^2=1$, and momenta along the $-\bm{x}+\bm{y}$ direction $k_{\bar{1}10}$.
In contrast to the case of Eq.~(\ref{eq:phM}), the antisymmetry and the $\mathbb{Z}_2$ number 
are defined in any fixed momenta $(k_{\bar{1}10}, k_z)$ in the Brillouin zone. 
Therefore, anywhere on the Fermi loop where the sign of $\mathbb{Z}_2$ number changes, 
the energy gap closes as shown in Fig.~\ref{fig:110surf}(a)-(e).

\subsection{Topology protected by non-symmorphic symmetry}
There is a topological property associated with non-symmorphic symmetry of iron-based superconductors, 
which is common to $B_{1u}$ and $B_{2u}$. 
In the combination of generators in Eq.~(\ref{eq:sym}), 
$C_4^2 P$ corresponds to glide mirror reflection, namely, translation by $\bm{t}=a\hat{\bm x}\!+\!a\hat{\bm y}$ followed by mirror reflection with respect to the $z=0$ plane.
In addition to glide mirror symmetry, this system hosts time-reversal and particle-hole symmetries Eqs.~(\ref{eq:PHS}) and (\ref{eq:Ts}). 
In this case, we can define a $\mathbb{Z}_4$ topological invariant protected by glide mirror symmetry~\cite{shiozaki2016}.  

We can evaluate this $\mathbb{Z}_4$ invariant based on the unification and subdivision of the topological phases associated with symmetry breaking and recovery. 
In general, in a system with glide mirror symmetry, one can recover mirror symmetry while retaining all the other symmetries.
In the presence of recovered mirror symmetry, the mirror Chern number $\nu_M=(\nu_{i}-\nu_{-i})/2$ is quantized to the $\mathbb{Z}$ number. 
Here, $\nu_{\pm i}= \sum_{E_n<0} \int dk_\parallel^2 \epsilon_{ij} (\partial_{k_i} \bm{u}_{n,\bm{k}_\parallel}^{\pm i\dag})(\partial_{k_j} \bm{u}_{n,\bm{k}_\parallel}^{\pm i})$ is the Chern number
for the eigenstate $H(\bm{k}_\parallel)\bm{u}_{n{\bm k}_\parallel}^{\pm i}=E_n\bm{u}^{\pm i}_{n{\bm k}_\parallel}$ with the mirror eigenvalue $M \bm{u}^{\pm i}_{n,\bm{k}_\parallel}=\pm i{\bm u}^{\pm i}_{n,\bm{k}_\parallel}$
on the mirror-symmetric momenta $D_M[\bm{k}_\parallel]=\bm{k}_\parallel$ and $H(\bm{k})$ is the one-particle or BdG Hamiltonian. 
If the recovery of symmetry can be achieved adiabatically, we have a correspondence of topological number between the systems with and without mirror symmetry 
shown in Table~\ref{table:corresp};
$\mathbb{Z}_2$ invariant in Eq.~(\ref{eq:z2}) for the normal state and $\mathbb{Z}_4$ invariant for the superconducting state
correspond to $\mathrm{mod}\ 2$ and $\mathrm{mod}\ 4$  parts of the mirror Chern number in the system with recovered mirror symmetry, respectively.

\begin{table}[t]
	\caption{Relations between the topological indices of the systems
	where the mirror symmetry is recovered and partially broken to the glide mirror symmetry.
	$\theta_n$ is the $\mathbb{Z}_n$ number with $\theta_n= 0$, $1$, $\cdots$, $n-1$. 
	$m$ is an integer.} \label{table:corresp}
	\renewcommand{\arraystretch}{1.5}
	\newcolumntype{C}{>{\centering\arraybackslash}X}
	\newcolumntype{P}[1]{>{\centering\arraybackslash}p{#1}}
	\begin{tabularx}{85mm}{  P{15mm} C C}
	\hline \hline
	$M_{\bm z}$ or $\tilde{M}_{\bm z}$ & Normal & Superconducting \\
	\hline
	Broken & $\theta_2=\nu_{M_{\bm z}} \mod 2$ & $\theta_4=\tilde{\nu}_{\tilde{M}_{\bm z}} \mod 4$ \\
	Recovered & $\nu_{M_{\bm z}}=\theta_2 + 2m$ & $\tilde{\nu}_{\tilde{M}_{\bm z}}=\theta_4 + 4m$ \\
		\hline
		\hline
	\end{tabularx} 
\end{table}

We apply this correspondence to the present Fe(Se,Te) system. 
By removing the displacement of chalcogen atoms by $b\rightarrow 0$ in Fig.~\ref{fig:lattice} or, in terms of the tight-binding model, setting the parameter $t_1^0\rightarrow 0$ and $t_{1,2}^+\rightarrow t_{1,2}^-$,
the mirror reflection symmetry with respect to the ${z=0}$ plane
\begin{eqnarray}
	M_{\bm z} = is_3 \sigma_0 \eta_0
\end{eqnarray}
is recovered. The energy gaps on the $k_z=0$ and $\pi$ planes do not close during this process. 
Applying the relation in Table~\ref{table:corresp} to the $\mathbb{Z}_2$ invariant $\theta_2(k_z)$ of the band structure in Fig.~\ref{fig:band}, 
the normal state with recovered mirror symmetry has
\begin{eqnarray}\label{eq:NmCh}
	\nu_{M_{\bm z}}(0) = 2m + 1, \quad \nu_{M_{\bm z}}(\pi) = 2m, 
\end{eqnarray}
with an integer $m$. 

In addition, the $B_{1u}$ and $B_{2u}$ gap functions are odd under the mirror reflection $M_{\bm z}\Delta M_{\bm z}^{-1}=-\Delta$.
In this case, the mirror symmetry operator for the BdG Hamiltonian is given as $\tilde{M}_{\bm z}=M_{\bm z} \tau_3$~\cite{ueno2013}. 
Within the weak pairing limit $\Delta\ll E_F$, the BdG Hamiltonian is $H\sim \mathrm{diag}(H_0,-H_0)$. 
Therefore, when the Fermi levels at the $\Gamma$ and $Z$ points are located in between the inverted bands as shown in Fig.~\ref{fig:band}, 
the mirror Chern number of the superconducting state can be evaluated as $\tilde{\nu}_{\tilde{M}_z}=2 \nu_{M_z}$ for $k_z=0$ and $\pi$. 
By substituting~Eq.~(\ref{eq:NmCh}), the $\mathbb{Z}_4$ invariant of the original system without mirror symmetry is evaluated as 
\begin{eqnarray}
	\theta_4(0) = 2, \quad \theta_4(\pi)=0. 
\end{eqnarray}
This indicates that the $B_{1u}$ and $B_{2u}$ superconductivity is non-trivial topological superconductivity protected by glide mirror symmetry.

We can observe the bulk-edge correspondence of this topological invariant
in the numerical solution for the $(110)$ surface (Fig.~\ref{fig:110surf}). 
The gapless states between $Y$ and $\Gamma$ in both the $B_{1u}$ and $B_{2u}$ states are
characteristics of the surface state protected by glide mirror symmetry.
The $B_{1u}$ and $B_{2u}$ gap functions are odd under the glide mirror reflection ${G}(\bm{k})=C^2_4 P e^{i\bm{k} \cdot\bm{t}}$. 
Hence, the symmetry of the BdG Hamiltonian is given as 
$\tilde{G}(\bm{k})H({\bm k})\tilde{G}^{-1}(\bm{k})=H(k_x,k_y,-k_z)$
with $\tilde{G}=G({k})\tau_3$. 
As $[\tilde{G},H(\bm{k})] = 0$ is satisfied on the $k_z=0$ plane, the eigenvalue of $\lambda_{\tilde{G}}(\bm{k})$ of $\tilde{G}(\bm{k})$ is a good quantum number. 
In addition, at the $\Gamma$ and $X$ points, time-reversal symmetry requires the Kramers degeneracy.
At the $\Gamma$ ($X$) point, as the glide mirror eigenvalue is $\lambda_{\tilde{G}}=\pm i (\pm 1)$, 
eigenstates with different (same) eigenvalues $\lambda_{\tilde{G}}$ form the Kramers pair.
In other words, between the $X$ and $\Gamma$ points, 
two Kramers pairs must exchange their eigenstates with different $\lambda_{\tilde{G}}$. 
This exchange typical for Mobius twisted surface state (hourglass fermion) for glide protected time-reversal 
invariant topological phases~\cite{shiozaki2016}.

\section{Conclusion}\label{sec:summary}
In summary, we have developed a theory of topological crystalline phases associated with the Dirac semimetallic band structure of iron-based superconductors. 
Based on the minimal tight-binding model, the Cooper pairing states between the $p$- and $d$-orbitals, 
which strongly mix with each other at the Fermi level, yield odd-parity superconductivity.
Moreover, these superconducting states have non-trivial topological invariants 
protected by the generators $P4/nmm$ space group of iron-based superconductors and 
hence exhibit characteristic Majorana flat surface states
and point and line nodes of the surface Fermi loop. 
The observed results show that iron-based superconductors are promising platforms to realize rich topological crystalline phases.

\begin{acknowledgments}
This work was supported by JSPS KAKENHI Grant numbers JP16K17755, JP17H02922, JP17J08855,
JSPS Core-to-Core program, and the Grants-in-Aid for Scientific Research on Innovative Areas ``Topological Material
Science,'' JSPS (Grant No. JP15H05855). 
This project was supported in part by JSPS and ISF under Japan-Israel Research Cooperative Program.
The numerical calculations were performed on XC40 at YITP in Kyoto University.
\end{acknowledgments}

%==========================================================================================================
\bibliography{reference}
\end{document}